\documentclass[%
 aip,
 amsmath,amssymb,
preprint,%
]{revtex4-1}

\usepackage{graphicx}
\usepackage{dcolumn}
\usepackage{bm}
\usepackage{float}

\usepackage[utf8]{inputenc}
\usepackage[T1]{fontenc}
\usepackage{mathptmx}
\usepackage{etoolbox}
\usepackage{xcolor}
\usepackage{soul} 

\makeatletter
\def\@email#1#2{%
 \endgroup
 \patchcmd{\titleblock@produce}
  {\frontmatter@RRAPformat}
  {\frontmatter@RRAPformat{\produce@RRAP{*#1\href{mailto:#2}{#2}}}\frontmatter@RRAPformat}
  {}{}
}%
\makeatother
\begin{document}
\title{Shearless bifurcations for two isochronous resonant perturbations}
\author{B. B. Leal}%
\affiliation{Institute of Physics, University of São Paulo, 05508-900 São Paulo, SP, Brazil}%
\author{M. J. Lazarotto}
\affiliation{Institute of Physics, University of São Paulo, 05508-900 São Paulo, SP, Brazil}%
\author{M. Mugnaine}
\affiliation{Institute of Physics, University of São Paulo, 05508-900 São Paulo, SP, Brazil}%
\author{A. M. Ozorio de Almeida}
\affiliation{Brazilian Center for Research in Physics, Rio de Janeiro, RJ, 22290-180, Brazil}
\author{R. L. Viana} 
\affiliation{Department of Physics, Federal University of Paran\'a, Curitiba, PR, 81531-980, Brazil}%
\author{I. L. Caldas} 
\affiliation{Institute of Physics, University of São Paulo, 05508-900 São Paulo, SP, Brazil}%

\date{\today}%

\begin{abstract}
In nontwist systems, primary shearless curves act as barriers to chaotic transport. Surprisingly, the onset of secondary shearless curves has been reported in a few twist systems. Meanwhile, we found that, in twist systems, the onset of these secondary shearless curves is a standard process that may appear as control parameters are varied in situations where there is resonant mode coupling. Namely, we analyze these shearless bifurcations in two-harmonic systems for the standard map, {the} Ullmann map, and {for the} Walker-Ford Hamiltonian flow. The onset of shearless curves is related to bifurcations of periodic points. Furthermore, depending on the bifurcation, these shearless curves can emerge alone or in pairs, and in some cases, deform into separatrices.
\end{abstract}

\maketitle
\begin{quotation}
Transport, in dynamical systems, is defined as the collective motion of many chaotic trajectories through the phase space. Regular trajectories can act as partial or total barriers, leading to a scenario of low or null transport, respectively. One example of a robust barrier is the shearless curve, a special solution for nontwist systems. Interestingly, such solutions can be found locally in twist systems, especially when there is a resonant mode coupling. In this work, we explore the possibility of local shearless curves in three different conservative systems. We find that  shearless curves can emerge in pairs or alone, depending on the bifurcation of the periodic points that they surround. The same scenario is observed in all systems studied, indicating that local shearless curves are a recurrent phenomenon in conservative twist systems.

\end{quotation}
    

\section{Introduction}
Non-integrable Hamiltonian systems are known for the coexistence of regular and chaotic solutions in the phase space. Due to the complexity involving the mixing of solutions with different characteristics, the movement of chaotic trajectories in the phase space can be partially or totally restricted by the regular structures which act as barriers, leading to a non-ergodic chaotic motion \cite{zaslavsky}. Area-preserving maps are useful tools for analyzing the behavior of these Hamiltonian systems. They can be specified by analytical discrete expressions or numerically obtained as Poincaré sections of successive intersections of the Hamiltonian flows\cite{lichtenberg}.

Among the conservative maps, we highlighted the class of \textit{twist maps}, composed by maps with a variational formulation that can be derived from generating functions \cite{mackay}. \textcolor{black}{The twist maps are known for satisfying the twist condition, \textit{i.e.}, in an angle-action portrait, as we vary the action, the iterated points will lie on different concentric circles with different time averages of the angle of rotation. }

\textcolor{black}{The time average rotation, represented by the rotation number, depends on the action.} For twist maps, \textcolor{black}{the dependency} is always monotonic, so the twist is always in the same direction \cite{reichl} and the angular displacement is in the same direction for all circles. However, it is possible to introduce area-preserving maps with a non-monotonic function. These maps are called \textit{nontwist}, since the twist condition is violated at some point in the phase space. With a non-monotonic rotation number, there is a circle \textcolor{black}{which} presents an extreme value and more than one circle presents the same rotation number.

Due to the violation of the twist condition (named non-degeneracy condition for Hamiltonian continuum systems)  different dynamical {phenomena} emerge in the phase space \cite{del1993,del1996}. Twin island chains and separatrix reconnection are examples of nontwist phenomenons which happens due to the \textcolor{black}{non-monotonicity of the rotation number~\cite{howard1984,delshams2000}. The solution in which the twist condition fails is called \textit{shearless/twistless curve}, since the derivative (frequently called the shear/twist) of the rotation number in relation to the action vanishes at such curve\cite{del1996}.} 

In general, an invariant curve can be regarded as a transport barrier. However, in twist maps, due to KAM theorem, these barriers are progressively destroyed as the perturbation strength is continuously increased. Shearless curves, on the other hand, are robust in the sense that they can survive the destruction of invariant curves on both sides of them (in the action variable). We shall use the word “transport barrier” for the shearless curves corresponding to local extrema of the rotation number profile.

The presence of shearless curves was identified in many physical systems, as the Rossby waves experiments in a rotating annular tank \cite{swinney1991,morrison1992}, in toroidal devices for plasma confinement as the TCABR\cite{marcus2008,marcus2008PoP} and Texas Helimak \cite{toufen2012} tokamaks, in mathematical models for Rossby waves in shear flows \cite{del1993}, magnetically confined plasmas \cite{caldas2012}, zonal flows in geophysical systems  and chaotic advection \cite{del2000,morrison2000}, to cite some examples.

The twist measures the rate of change of frequencies of different invariant curves \cite{delshams2000} and for twist (nontwist) maps, its sign is constant (changes). These invariant circles are centered in the origin and the rotation number is a global measure, since it is related to the rotation of the points around the center of the invariant circles \cite{reichl}. In the last decades, the concept of rotation number was generalized, and it is possible to define its value for the rotation of points around fixed points in the phase space. Therefore, the concept of \textit{internal rotation number} was proposed as the measure of torsion of each torus with respect to its elliptic point \cite{dullin2000,abud2012,abud2014}.

Along with the idea of internal rotation number, the possibility of shearless (twistless) curves was also demonstrated in twist maps. In this case, the shearless curve is secondary and exhibits an extreme value for the internal rotation number. Dullin, Meiss and Sterling\cite{dullin2000} observed the existence of twistless curves in the neighborhood of an elliptic point which goes through a tripling bifurcation. Abud and Caldas\cite{abud2012} also observed these twistless curves in the standard twist map in the neighborhood of tripling and quadrupling bifurcations. They also observed such curves in twist maps for field lines in tokamaks \cite{abud2014}. Secondary shearless curves can also be observed in nontwist maps \cite{abud2012}.

According to Dullin, Meiss and Sterling, the presence of shearless torus on the neighborhood  of a periodic orbit that goes through a tripling bifurcation is generic, and the authors also affirm that the twist of an orbit can be forced to be zero if there is a sufficient quantity of parameters \cite{dullin2000}. Interestingly, secondary shearless curves were also identified in the standard twist map \cite{abud2012}, models of optical lattices \cite{lazarotto2024} and of large aspect ratio tokamaks with ergodic limiter \cite{leal2024}. With these results, we suspect that secondary shearless curves are more common than was first assumed. In order to investigate the generality of secondary shearless curves, we propose to study different dynamical system, investigating the possibility of these curves around different types of bifurcations.

In this paper, we study the emergence of secondary shearless curves in three different twist systems. The first system is \textcolor{black}{a two-parameter twist} map, known as the two-harmonic standard map, presented in Ref. \cite{mugnaine2024}, composed by the addition of a second resonant perturbation in a generic version of the standard map. This map can be considered as a model for the competition of different resonant modes and the resulting isochronous islands go through different bifurcation, such as pitchfork and saddle-node. The second system is presented in Ref. \cite{leal2024}, a magnetic field line map, adapted from the Ulmann map \cite{ullmann2000}, composed of two parts: the first dictating the evolution of the field lines between two rings of the ergodic limiter, and the second which describes the action of the limiter as an impulsive perturbation. The last studied system is the well-established Walker-Ford Hamiltonian flow \cite{Walker-Ford}, which describes the effect of resonances on the appearance of stability islands. Our investigations of such systems is based on the computation of the internal rotation number profile and the phase space analysis of the extreme points in the profile and the presence of such curves around elliptic points.

Our conclusion is that secondary twistless tori are more common in twist systems than we expected. All three studied systems present such torus for different configurations and sets of parameters. As we observe, the emergence of shearless curves are related to the bifurcations of periodic points. Furthermore, these shearless curves can emerge alone or in pairs, depending on the bifurcation it is related to.

The present paper is organized as follows.  Section \ref{sec:SM2m} is dedicated to the investigations around the two-harmonic standard map, the simplest system analyzed, which will give us a simpler explanation about the emergence of secondary shearless torus. In Section \ref{sec:UM}, we present the results on the modified Ulmann map and the impact of different arrangements of resonant modes in the rise of shearless curves. The results on the Walker-Ford Hamiltonian flow  are presented in Section \ref{sec:WF}, where, in contrast with previous sections, we study the emergence of shearless torus in time-continuum system. Our conclusions are presented in Section \ref{sec:Conclusions}.

\section{Two-harmonic standard map and local shearless tori}
\label{sec:SM2m}
The two-harmonic standard map, also called standard map with two modes, is defined by the equations,
\begin{eqnarray}
    \begin{aligned}
    y_{n+1}&=y_n-\dfrac{K_1}{2\pi m_1} \sin(2 \pi m_1 x_n)-\dfrac{K_2}{2 \pi m_2} \sin(2 \pi m_2 x_n),\\
    x_{n+1}&=x_n+y_{n+1},        
    \end{aligned}
    \label{sm2m}
\end{eqnarray}
where $x$ and $y$ are taken mod 1, $K_1$ and $K_2$ are the perturbation amplitudes, and $m_1$ and $m_2$ are the modes of the resonant perturbations. Here, we consider $K_1, K_2 \in \mathbb{R}^*_+$ and $m_1,m_2 \in \mathbb{N}$. The two-harmonic map is a generalization of the extended standard map (ESM), where $m_1=1$ and $m_2=2$. The ESM can be obtained in the study of a one-dimensional lattice of particles which interact elastically with the nearest neighbor \cite{greene1987,johannesson1988}. 

The map (\ref{sm2m}) was proposed as a model for the competition between two resonant modes $m_1$ and $m_2$ \cite{mugnaine2024}.  The two resonant perturbations generate different quantities of islands in the same region of the phase space\cite{Sousa}. As it was investigated in Ref. \cite{mugnaine2024}, the number of isochronous islands in the line $y=0$ is equal to the number of the mode of the resonant term. Thus, if we observe $m_1$ ($m_2$) islands in $y=0$ in the phase space of map (\ref{sm2m}), we say the mode $m_1$ ($m_2$) is predominant. The transitions from the chain of islands related to the mode $m_1$ to the chain of island of the second mode $m_2$ can happen with or without intermediate modes. For more information about the transition routes, see Ref. \cite{mugnaine2024}.

The islands in the line $y=0$ present period and frequency equal to unity for both modes $m_1$ and $m_2$. The global winding number is defined by the limit
\begin{eqnarray}
    \omega = \lim_{n\to \infty} \dfrac{x_n-x_0}{n}
    \label{wn}
\end{eqnarray}
and it is also equal to unity for these islands. The two resonant perturbations can act in the same winding number surface, and the quantity of chains of islands in this surface varies according to the perturbation parameters $m_{1,2}$ and $K_{1,2}$. It is worth noting that to determine (\ref{wn}), $x_n$ must be computed  without taking its mod.

\begin{figure}[H]
    \centering
    \includegraphics[width=0.5\textwidth]{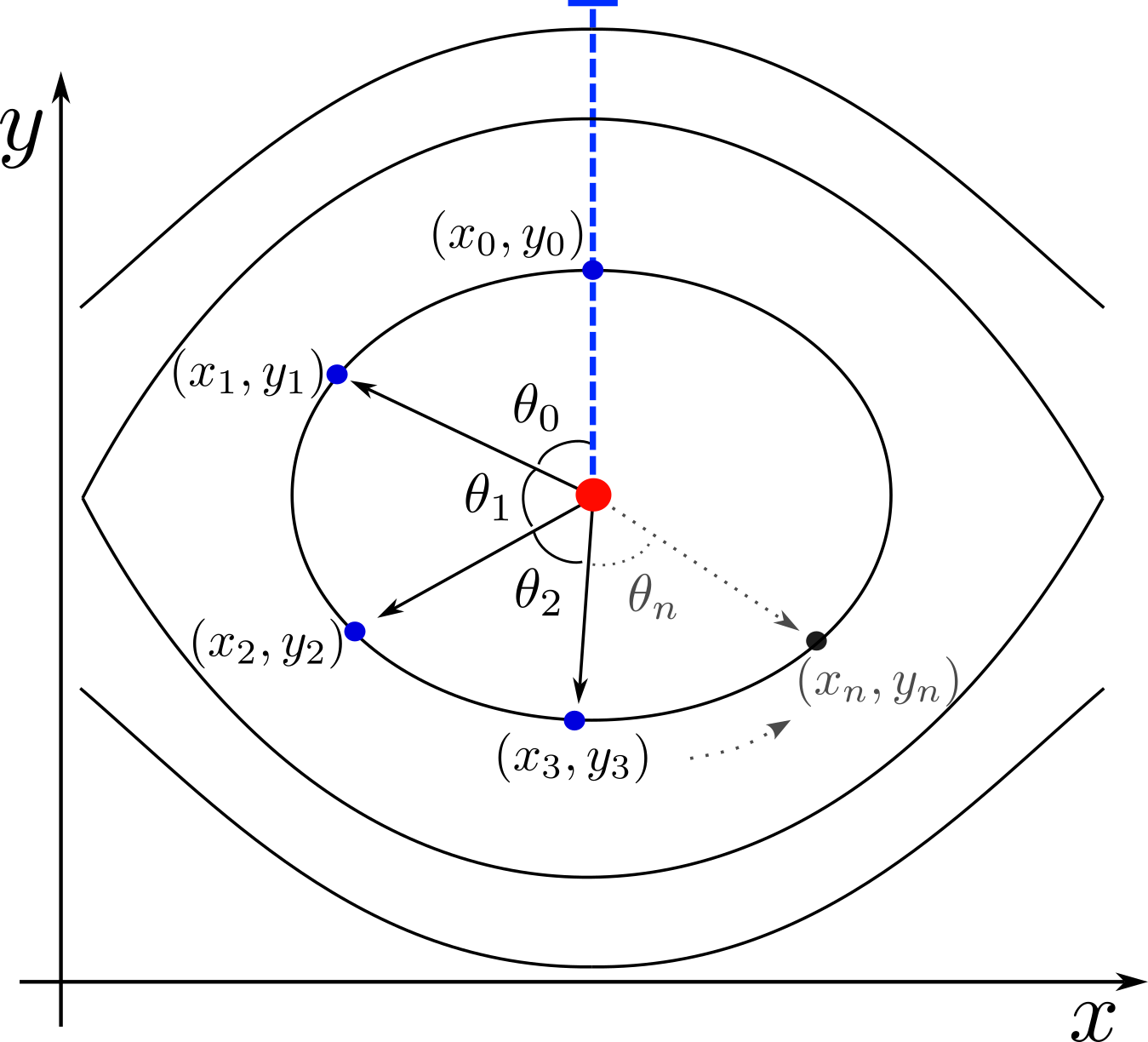}
    \caption{Schematic figure for the calculation of $\omega_{in}$. For an initial point $(x_0, y_0)$ over the winding profile reference line (dashed blue line),  the inner angle $\theta_n$ is sequentially evaluated relative to the center of the island. }
    \label{fig:winding_method}
\end{figure}

The standard map is a twist map, \textit{i.e.,} it satisfies the twist condition $\frac{\partial x_{n+1}}{\partial y_n} \ne 0$ for every point in the phase space. As a consequence, if we compute the winding number profile for different orbits with initial conditions on a fixed line $x$, the value of $\omega$ changes monotonically with $y$ and forms a plateau in regions where there are islands. The addition of a second resonant perturbation in the standard map does not change the twist property of the system, the winding number profile remains monotonic. However, non-monotonic profiles can be obtained if we compute the rotation number locally by the measure of the rotation  of a single island with respect to its elliptic point \cite{Abud,abud2012,leal2024}. Thus, we can define a internal rotation number $\omega_{in}$, which is computed by the equation \cite{abud2012},
\begin{eqnarray}
    \omega_{in}=\lim_{n\to\infty} \dfrac{1}{2\pi n} \sum_{n=1}^{\infty} P_n\hat{\theta}P_{n+1}
    \label{win}
\end{eqnarray}
with $\theta=P_n\hat{\theta}P_{n+1}$ being the angle between two consecutive points in the phase space. An schematic image about the computation of $\omega_{in}$ is shown in Fig. \ref{fig:winding_method}. Due to the normalization with $2\pi$, the internal rotation number is in the range $[0,1]$. Just as for the global correspondent $\omega$, if $\omega_{in}$ assumes a rational or irrational value, it means the orbit is periodic or quasi periodic, respectively. The internal rotation number also does not converge for chaotic solutions.

The non-monoticity of the internal winding number is characterized by an extreme point in the profile, and such point refers to the local shearless curve, also called twistless torus \cite{Abud, abud2012,leal2024,dullin2000}. The shear, or twist, are related to the derivative of the rotation number with respect to the action \cite{dullin2000}, so in an extreme point of $\omega_{in}$ we have $\omega'_{in}=0$ and the lack of shear or twist.

The system (\ref{sm2m}) is a simple model for twist Hamiltonian systems with multiple resonant perturbations. Following the analysis performed in \cite{leal2024}, we investigate the existence of twistless tori for different combinations of modes $m_1$ and $m_2$ and in different transition routes discussed in Ref. \cite{mugnaine2024}.

\subsection{Emergence of one shearless curve}
\label{sec:A}
\begin{figure}[!h]
	\begin{center}
		\includegraphics[width=1.0\textwidth]{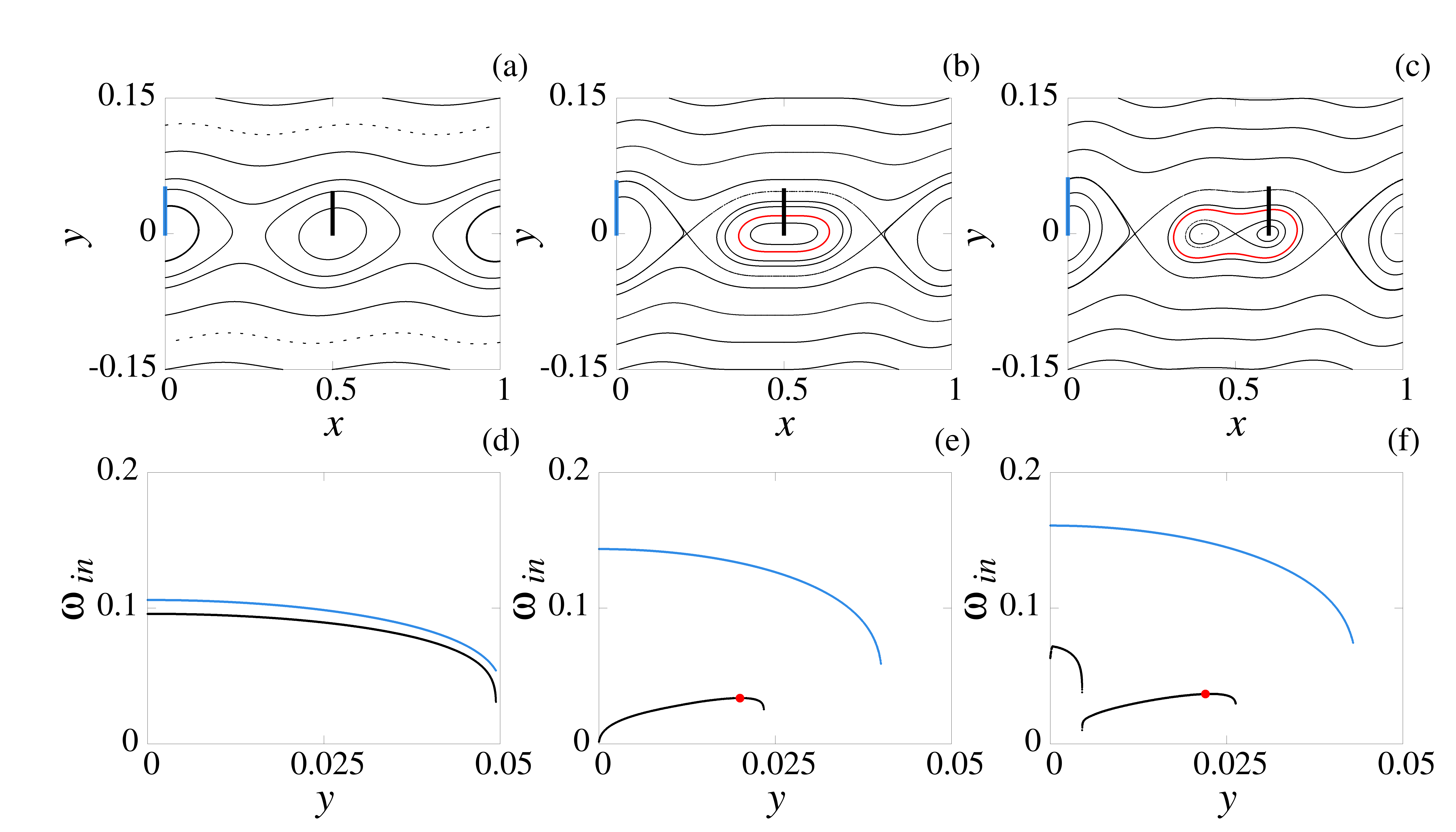}		
		\caption{Transition from mode $m_1=2$ to mode $m_2=3$ in the two-harmonic standard map. For all cases, $K_1=0.1$. In (a) and (b), we have the predominance of mode $m_1=2$, with $K_2=0.01$ and $K_2=0.1$, respectively. For $K_2=0.15$, we have the phase shown in (c), where the pitchfork bifurcation has already happen and we observe three elliptic points. The internal winding number profile for each phase space (a)-(c) are shown in panel (d)-(f), respectively. The profiles are computed in the blue and black lines present in the phase space, and each profile correspond to the line with the same color. The red curves and red points indicate the shearless torus.}
		\label{fig1}
	\end{center}
\end{figure}

As was shown in Ref. \cite{mugnaine2024}, one of the possible transitions from one chain of islands to the other is by pitchfork bifurcations. As an example, we take the transition $m_1=2 \to m_2 =3$, where the elliptic point at $(x,y)=(0.5,0)$ goes through a pitchfork bifurcation, it is replaced by a hyperbolic point and other two elliptic points emerge. This transition is presented in Fig. \ref{fig1}. Along with the phase spaces, we compute the internal winding number for each island in the phase space in relation to their elliptic points. The profiles are also presented in Fig. \ref{fig1}.

As shown in Fig. \ref{fig1}, as the value of $K_2$ increases, the island of period one centered in the point $(x,y)=(0.5,0)$ changes its form, as it can be seen by the comparison between panel (a) and (b). The shape of the island around point $(x,y)=(0,0)$ remains the same. If we increase $K_2$ more, we have the phase space shown in Fig. \ref{fig1} (c), where a pitchfork bifurcation happened and a hyperbolic point replaces the stable point and two others elliptic points emerge in line $y=0$. In this case, the transition from mode $m_1$ to $m_2$ happens by a single pitchfork bifurcation.

The internal winding number profiles presented in Fig. \ref{fig1} (d)-\ref{fig1}(f) were calculated with $5. 10^3$ initial conditions, distributed in the blue and black lines shown in the phase space, iterated by $10^4$ iterations. If the limit (\ref{win}) converges, we plot the value of $\omega_{in}$ for the respective value of $y$. In Fig. \ref{fig1} (d), we observe both profiles have a monotonic behavior and decrease as the value of $y$ increases, thus no internal shearless torus is present in the phase space. The same behavior repeats for the island centered in $(x=0,y=0)$ for greater values of $K_2$, as it is shown by the blue curves in Fig. \ref{fig1} (e) and \ref{fig1}(f).

Differently, shearless torus can exist around the elliptic point at $(x,y)=(0.5,0)$, since extreme points can be observed in the $\omega_{in}$ profiles in Fig. \ref{fig1} (e) and \ref{fig1}(f). These extreme points are marked by the red circles and the respective shearless tori are also marked in red in phase spaces of Fig. \ref{fig1} (b) and \ref{fig1} (c). The shearless tori are identified by maximum points in the the $\omega_{in}$ profiles and the torus is in the middle region of the islands in the phase spaces, it emerges before the pitchfork bifurcation [Fig. \ref{fig1} (b)] and it persists after it [Fig. \ref{fig1} (c)].

\subsection{Emergence of multiple shearless curves}
\label{sec:B}
\begin{figure}[!h]
	\begin{center}
		\includegraphics[width=1.0\textwidth]{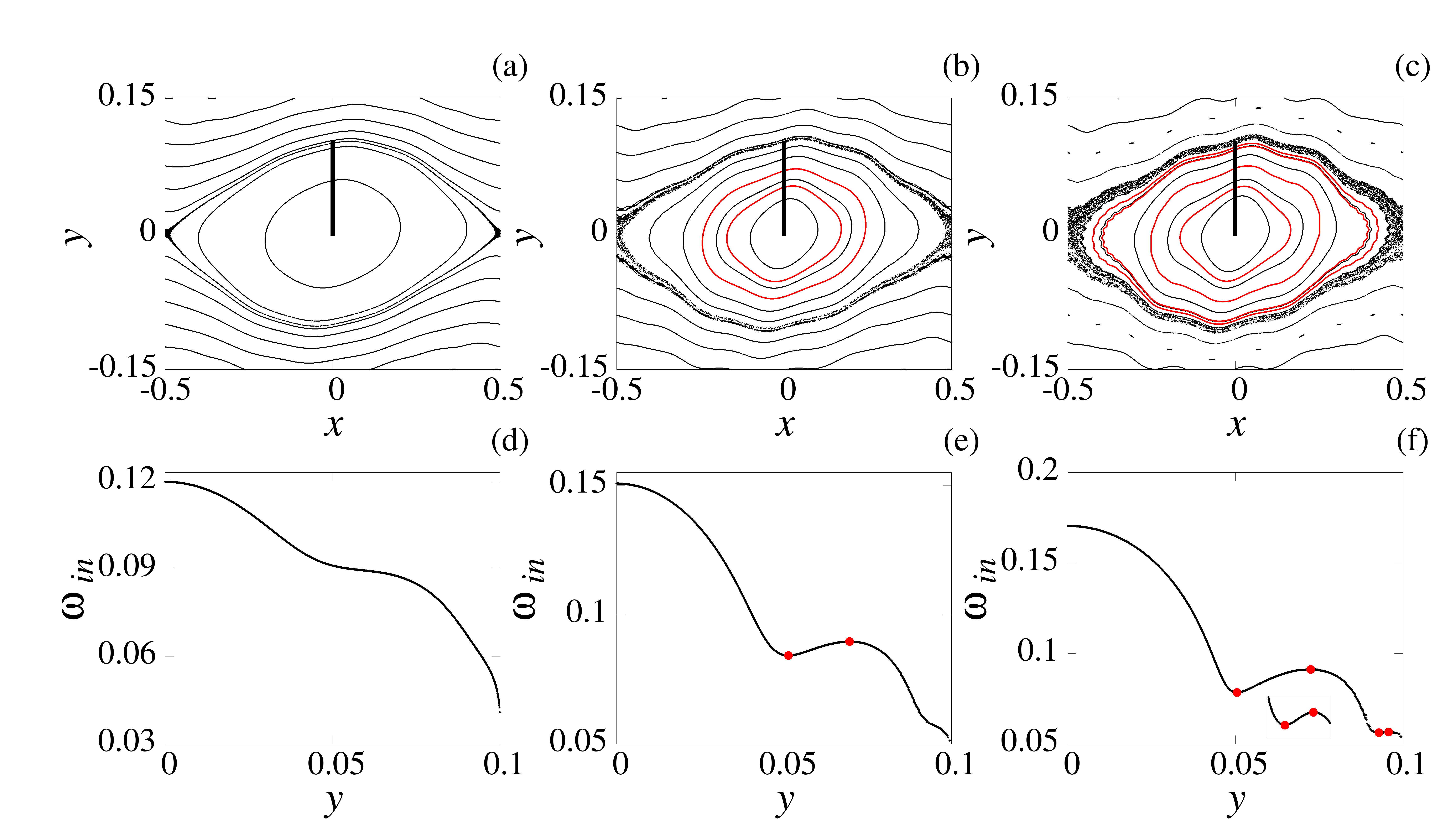}		
		\caption{Effect of the second mode $m_2=5$ in the first mode $m_1=1$ for $K_1=0.1$ and different amplitudes $K_2$. In (a) $K_2=0.04$, (b) $K_2=0.12$, (c) $K_2=0.18$. For all panels, the first mode is predominant, since only one elliptic point and is in the line $y=0$. The action of the second mode is more noticeable in panel (b) and (c) where we observed the distortion of the islands. The red curves indicate the shearless tori. The internal winding number profiles respective to phase spaces (a)-(c) are shown in panels (d)-(e) respectively, with the extremes highlighted by the red circles which indicate  the position of the shearless curves.}
		\label{fig2}
	\end{center}
\end{figure}

Besides the pitchfork bifurcation, saddle-node bifurcations can also compose the transition routes from one mode to the other. It was observed that saddle-node bifurcations happen in pairs inside islands which, due to the emergence of new periodic points, change its form. As an example to study the possibility of shearless tori in these scenario, we examine the transition from mode $m_1=1$ to mode $m_2=5$. Repeating the analysis performed in the last section, the phase spaces and the internal winding number profiles for these modes are shown in Fig. \ref{fig2}.

In the phase spaces of Fig. \ref{fig2}, we observe the predominance of the mode $m_1=1$ for all the amplitudes $K_1$ and $K_2$ studied. We do not observe the five islands related to mode $m_2=5$ but we can identify the effect of the second mode by the distortion on the island of Fig. \ref{fig2} (b) and \ref{fig2} (c). However, from the results shown in Ref. \cite{mugnaine2024}, we know that  four islands will emerge inside the big distorted island with center in $(x,y)=(0,0)$ by four simultaneously saddle-node bifurcations.

In panel (d)-(f) of Fig. \ref{fig2}, we present the respective internal winding number profiles computed in the black line $x=0$ shown in the phase spaces of Fig. \ref{fig2} (a)-\ref{fig2} (c). Just as the first scenario presented in Fig. \ref{fig1} (d), the $\omega_{in}$ profiles also decreases monotonically in Fig. \ref{fig2} (d). However, the function related to these profiles are different between each other: while for Fig. \ref{fig1} (d) the profile resembles half of a parabola, in Fig. \ref{fig2} (d), the function is similar to a sum of $\sin(x)$ with different frequencies. Increasing the amplitude $K_2$ of the second mode, we have the profile shown in Fig. \ref{fig2} (e) with two extreme points, one maximum and one minimum, highlighted by the red points. These two points indicate the presence of two shearless curves in the phase space and such curves are highlighted in red in Fig. \ref{fig2} (b). In this panel (e), for $y$ close to $y=0.1$, the profile present the same shape of the profile in panel (d), indicating some aspect of similarity.

In the last panel, Fig. \ref{fig2} (f), we have the $\omega_{in}$ profile for the islands of phase space in Fig. \ref{fig2} (c). This case represents the largest studied value of $K_2$. From the phase space, we can identify four extreme points, two maxima and two minima, where one pair is the same pair of Fig. \ref{fig2} (e) and the new pair emerges close to $y=0.1$ and is zoomed in the inset. The four respective twistless curves are shown in red in Fig. \ref{fig2} (c).

From the sequence of internal winding number profiles shown in Fig. \ref{fig2} (d)-\ref{fig2} (f), we can identify a pattern on the modification of the profile and the emergence of shearless curves. First, the profile is monotonic and presents different concavities. As we increase the amplitude of the second mode, an inflection point is present and thereafter a pair of maximum-minimum points emerge in the profile, indicating the emergence of a pair of shearless curves. This sequence can happen multiple times in the profile and each inflection point will give rise to two shearless curves.

\subsection{Emergence of shearless curves in different islands}
\label{sec:2}
\begin{figure}[!h]
	\begin{center}
		\includegraphics[width=1.0\textwidth]{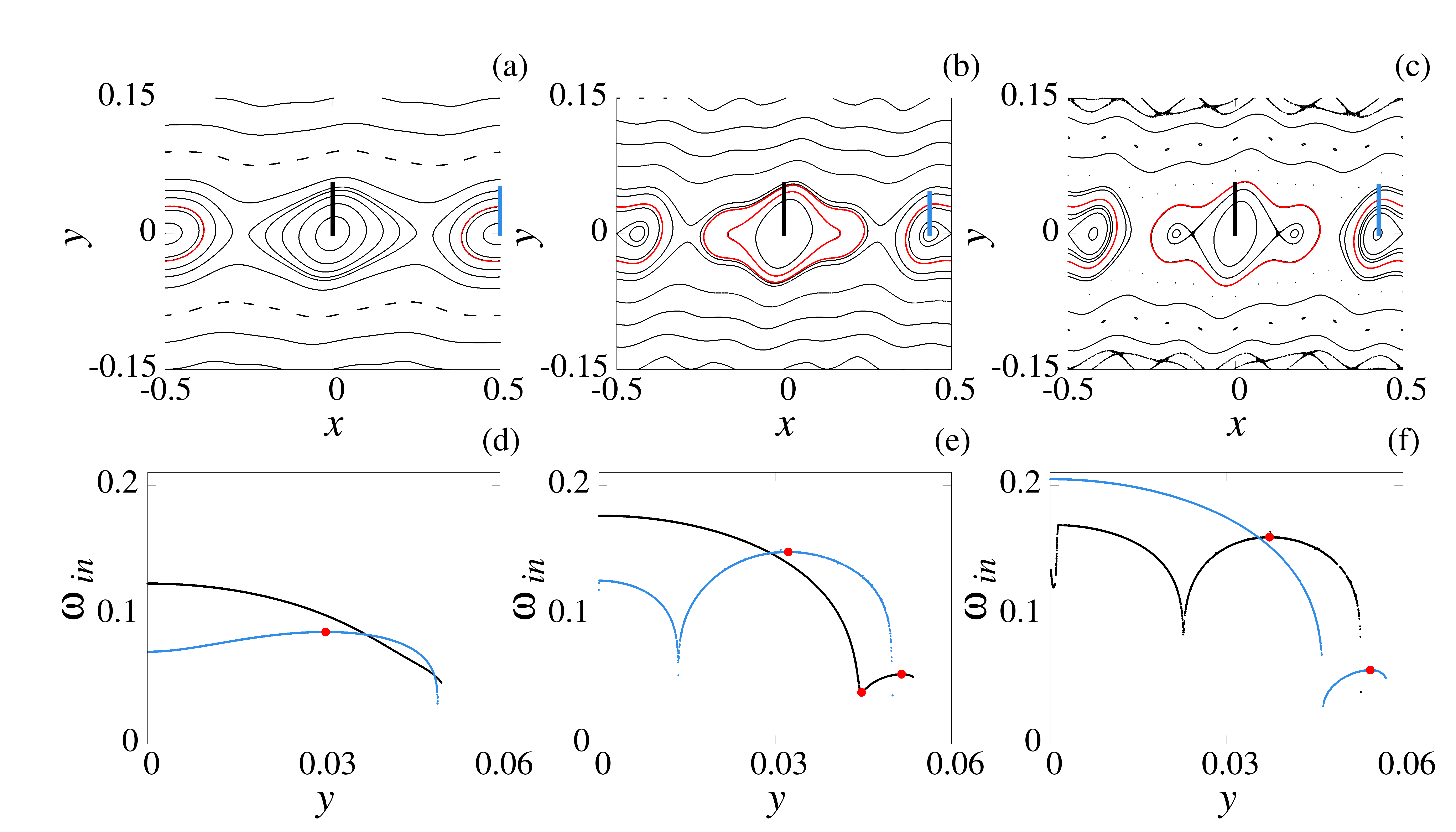}		
		\caption{Emergence of twistless torus in different islands. For all panels $K_1=0.1$ and the amplitude of the second mode is (a) $K_2=0.05$ (b) $K_2=0.2$, (c) $K_2=0.3$. While we observe the emergence of just one shearless torus around the elliptic point at $(x,y)=(-0.5,0)$, we observe two curves around the point $(x,y)=(0,0)$}
		\label{fig3}
	\end{center}
\end{figure}

Different isochronous islands can go through different bifurcations as one parameter changes. Here, we investigate if it is possible that shearless curves emerge in different islands. For this to happen, bifurcations must happen inside the islands. Thus, we choose the configuration $m_1=2$ and $m_2=5$, where a pitchfork and saddle-node bifurcations happen inside the two initial islands. The transition from mode $m=2$ to $m=5$ is shown in Fig. \ref{fig3} (a)-\ref{fig3} (c). This transition presents an intermediate mode, $m=3$, due to the pitchfork bifurcation that happens first in the island on $x= 0.5$, as we can observe in Fig. \ref{fig3} (b). After this first bifurcation, two elliptic points emerge inside the island centered in $x=0$ by two saddle-node bifurcations, as shown in Fig. \ref{fig3} (c).

The internal winding number profiles calculated in relation to the two initial elliptic points are shown in Fig. \ref{fig3} (d). The black curve, related to the elliptic point at $(x,y)=(0,0$) monotonically decreases while the blue curve, related to the point at $(x,y)=(0.5,0)$ exhibit a maximum point, indicating the presence of a shearless curve. 

If we increase the parameter $K_2$ from $K_2=0.05$, of Fig. \ref{fig3} (d), to $K_2=0.2$, we have the $\omega_{in}$ profiles of Fig.\ref{fig3} (e). In this case, both profiles shown extreme points, indicated by the red circles, and each profile have a different behavior. For the island centered in $x=0$, we observe the emergence of a pair of shearless curves since the $\omega_{in}$ profile exhibit one maximum and one minimum point, just as it was discussed in Section \ref{sec:B}.  Otherwise, the blue profile, computed in relation to the elliptic point which emerge in the pitchfork bifurcation, present a maximum value, indicated by the red circle, and one cusp indicating the separatrix which pass through $x=0.5$ inside the island. 

For $K_2=0.3$, we have the profiles shown in Fig. \ref{fig3} (f). In this case, one of the shearless curves around $x=0$ is already broken and both profiles present a cusp and a maximum value, indicated by the red circles. Both cusps represent separatrices. Inside the pair of shearless curves emerge two pairs of elliptic-hyperbolic points by saddle-node bifurcations.

From our observations, the emergence of shearless curves precedes the bifurcations which occur in the periodic points. Besides the shearless curves shown in this manuscript, we analyze the emergence of these curves for all combinations of $m_i=1, 2, 3, 4, 5$ and 6 for $i=1$ and 2, and we notice that the emergence of just one shearless curve happens before a pitchfork bifurcation while the emergence of a pair precedes saddle-node bifurcations.

\section{Ullmann's map double coupling}
\label{sec:UM}

For the second studied system, we have a symplectic mapping \cite{ott2002book} that describes the evolution of a magnetic field line configuration capable of confining a plasma along a torus that is periodically perturbed by an ergodic magnetic limiter, which is essentially  a set of coils through which electric currents flow. This mapping employs coordinates radial $r$ and poloidal $\theta$ to describe the position of a magnetic field line (analogous to cylindrical coordinates). A more detailed discussion about the relation between the map and the physical system is presented in Appendix \ref{apend}

The equations that describe the magnetic field  line positions, at the plasma confinement equilibrium, are given by (\ref{UllmannEqulRadial}) and (\ref{UllmannEqulPoloidal})  

\begin{equation}
    r^* = r,
    \label{UllmannEqulRadial}
\end{equation}

\begin{equation}
    \theta^* = \theta + \frac{2 \pi}{N q(r^*)},
    \label{UllmannEqulPoloidal}
\end{equation}
where $N$ is the number of coils and $q(r)$ is the cylindrical safety factor \cite{wesson2004book} determined by the poloidal and toroidal magnetic fields. We adopted the same safety factor as in reference\cite{ullmann2000}.

The equations (\ref{UllmannEqulRadial}) and (\ref{UllmannEqulPoloidal}) determine the toroidal evolution of a magnetic field line that has been previously localized in $(r, \theta)$ to the position $(r^*, \theta^*)$, after $2 \pi / N$ rad toroidally shift, and before the perturbation due to the ergodic magnetic limiter. The ergodic limiter is formed by two pairs of set of coils and each one produce a mode perturbation of type $(m, n)$ and, between coils of the same pair, there is a poloidal twist generated by $\alpha_i = \pi n_i / m_i$, where $n_i$ is the toroidal number of the perturbation $(m_i, n_i)$ chosen by us.

The equation
\begin{equation}
    r^* = r + \frac{b m_i C_i}{m_i - 1} \left( \frac{r}{b} \right)^{m_i - 1} \sin[m_i(\theta^* + (j_i - 1) \alpha_i)]
    \label{ullmannPertRad}
\end{equation}
determines how the radial position $r^*$ is affected by the coil $j_i$, $j = 1, 2$, of the pair $i$, $i = 1, 2$; and the coordinate $r$ is the field line position after the perturbation. It should be noted that (\ref{ullmannPertRad}) determines the value of $r$ implicitly. Therefore, in order to determine it, some numerical method able to find roots of polynomials is necessary. Here we used the Newton-Rapshon  method.

The equation that describes the $\theta$ change due to the perturbation related to the ergodic limiter is
\begin{equation}
    \theta = \theta^* - C_i \left( \frac{r}{b} \right)^{m_i - 2} \cos{[m_i(\theta^* + (j_i - 1) \alpha_i)]},
    \label{ullmannPertPol}
\end{equation} 
where $\theta^*$, {as commented before}, is the poloidal field line position immediately before the perturbation coil, while $\theta$ is the poloidal position after it. The parameter $C_i$ is a {dimensionless} constant,
\begin{equation}
    C_i = \frac{2 \epsilon_i m_i g a^2}{q(a) R_0 b^2},
\end{equation}
proportional to the rate $\epsilon_i$ of electric current in each pair $i$ and the plasma current $I_p$, $\epsilon_i = I_i / I_p$, the length $g$ of the coils, the square $a^2$ of the column plasma radio and inversely proportional to the major radio $R_0$.

The ergodic magnetic limiter configuration that we have adopted compels the Ullmann map to be a composition of equations of the equilibrium and the perturbation calculated four times. For a single complete revolution about the torus performed by a magnetic field line, we must compute equations (\ref{UllmannEqulRadial}), (\ref{UllmannEqulPoloidal}), (\ref{ullmannPertRad}) and, (\ref{ullmannPertPol}), in this sequence, four times. Thus, it follows that there is no explicit formula from $(r_n, \theta_n)$, the $n$-th intersection of the field line with the  Poincaré section, to the subsequent intersection $(r_{n+1}, \theta_{n+1})$

We have chosen a setting of coils in such a way that the same magnetic surface is perturbed with two different resonant modes $(m_1, n_1)$ and $(m_2, n_2)$. The ratio $m/n$ will perturb the magnetic field line with winding number $\omega = n/m$. All the results below are examples of perturbation in the magnetic field line surface with rotational number $1/3$ {and for a better visualization of the Poincaré Sections we adopted the transformation of coordinates $(X = \theta / 2 \pi, Y = (b - r) / b)$}.

\subsection{Emergence of one shearless curve}

Initially, we investigate the arrangement $\{(3, 1), (12, 4)\}$. The set of coils responsible for resonances of the type $(3, 1)$ is maintained constant at the value $\epsilon_1 = 0.005648$, while the second parameter, $\epsilon_2$, is changed. The magnitude of this second parameter determines whether or not chains of islands of type $(12, 4)$ emerge.

We begin the investigation with $\epsilon_2 = 0.000113$. In this situation, there is only the island chain $(3, 1)$. We follow one of the islands compounding this mode, the island whose $x$ coordinate falls within the interval $[0.2, 0.6]$. As long as the value of $\epsilon_2$ increases, bifurcations will occur, and for a certain value of $\epsilon_2$  onward the system will be in the mode $(12, 4)$. It should be noted that each mode is counted according to the number of stable (or unstable) periodic orbits of period $3$. 

\begin{figure}[H]
	\begin{center}
		\includegraphics[width=1.0\textwidth]{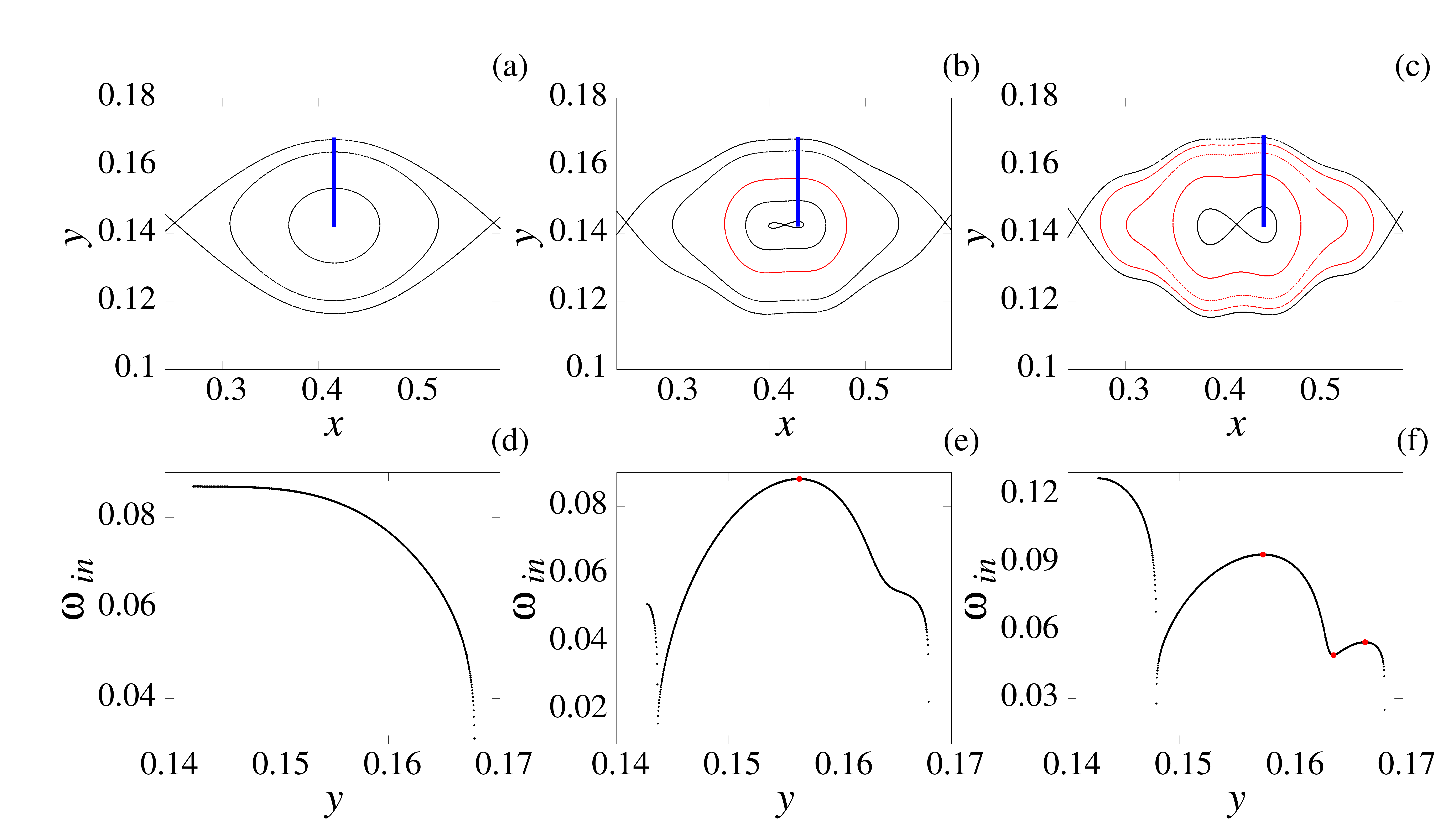}		
		\caption{Arrangement $\{(3, 1), (12, 4)\}$. The panels (a), (b), and (c) show respectively the same island for $\epsilon_2 = 0.000113,~ 0.001130$ and $0.004519$. For each one of these cases, there is the internal winding number profile in the panels (d), (e) and (f). In all the cases $\epsilon_1 = 0, 005648$.}
		\label{arrangment3_1_12_4}
	\end{center}
\end{figure}

At Fig. \ref{arrangment3_1_12_4} (d), there is the winding number profile related to the island with perturbation $\epsilon_2 = 0.000113$ of Fig. \ref{arrangment3_1_12_4} (a). The blue line segment is comprised of approximately $10^3$ points, each representing an initial condition that contributes to the formation of the winding number profile.

The initial profile in Fig. \ref{arrangment3_1_12_4} (d), is monotonic. Whereas the subsequent case, represented in Fig. \ref{arrangment3_1_12_4} (e), exhibits a nonmonotonic profile, there are two points of local extremes, indicated by highlighted points. However only one of these points defines a shearless curve that is represented in red in Fig. \ref{arrangment3_1_12_4} (b). The other local extreme, one local minimum, determines a separatrix that surges with the saddle-node bifurcation that occurs at the center of the island, as shown in Fig. \ref{arrangment3_1_12_4} (b). At a first glance, it appears that the type of bifurcation that has occurred is a pitchfork: the elliptical point in Fig. \ref{arrangment3_1_12_4} (a) has become a hyperbolic point in Fig. \ref{arrangment3_1_12_4} (b) and two elliptical points have simultaneously raised. However, what is actually occurring is that the elliptical point of  Fig. \ref{arrangment3_1_12_4} (a) is forced to drift to the right, and then a saddle-node bifurcation occurs. This is the reason that our blue line segment is shifted to the right when compared to Fig. \ref{arrangment3_1_12_4} (a).
\subsection{Emergence of multiple shearless curves}

Now, we are coupling the mode $(3, 1)$ with mode $(15, 5)$. When the two modes are coupled and $\epsilon_2$ is varied, four bifurcations occur before the island chain $(15, 5)$ be formed. As $\epsilon_2$ increases, the system will pass, by saddle-node bifurcations, through the modes $(6, 2)$, $(9, 3)$, $(12, 4)$ until it reaches the $(15, 5)$ one. Each one of these modes can be related to a new pair of shearless curves.

\begin{figure}[H]
	\begin{center}
		\includegraphics[width=0.6\textwidth]{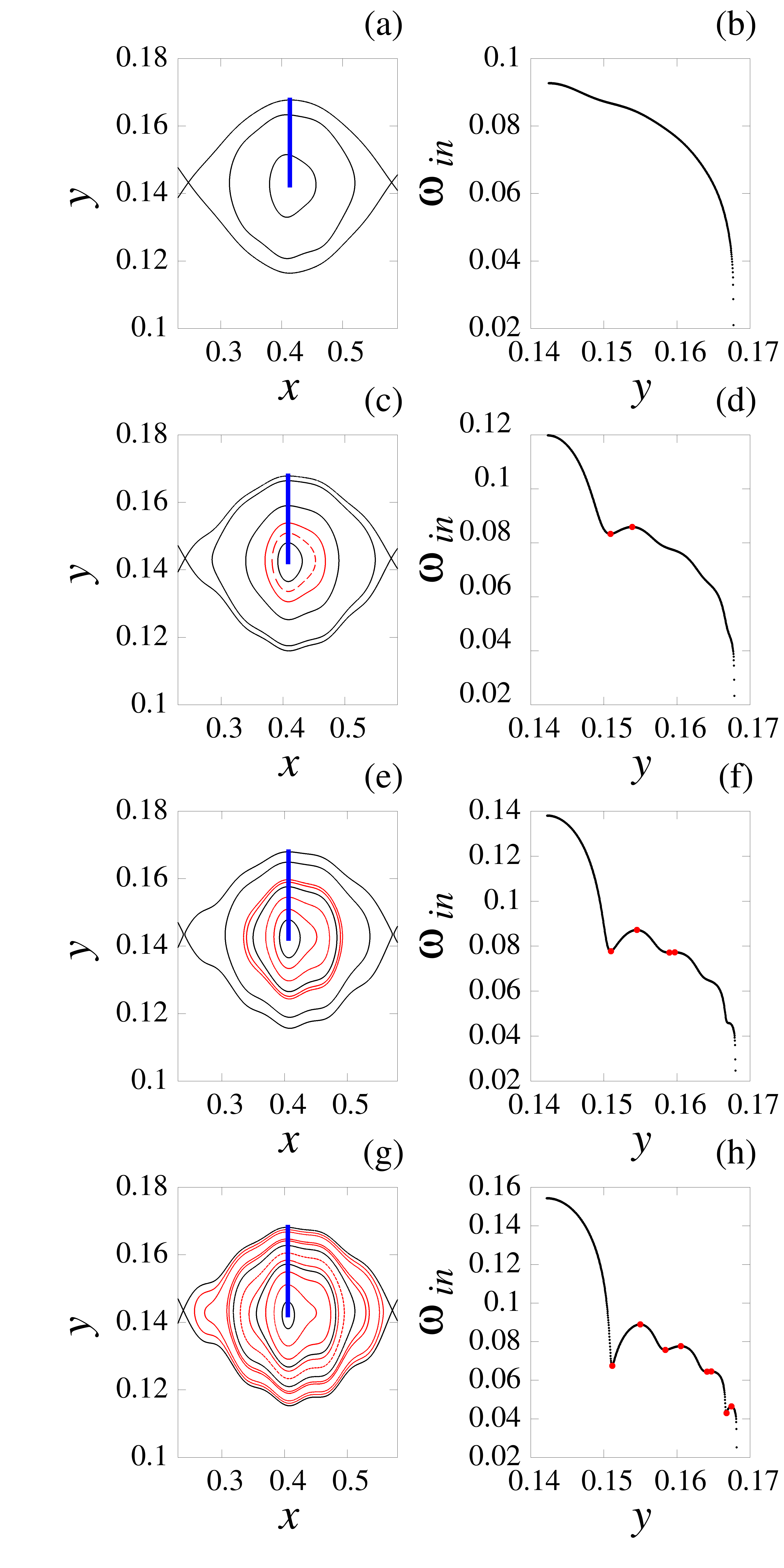}		
		\caption{Arrangement $\{(3, 1), (15, 5)\}$. The internal winding number profile, figs.(b, d, f) and (h), is determined for the same island: (a) $\epsilon_2 = 0, 000113$, (c) $\epsilon_2 = 0, 002259 $ (e) $ \epsilon_2 = 0, 003389 $ and (f) $\epsilon_2 = 0, 004519$. In all the cases $\epsilon_2 = 0,005648.$}
		\label{UllmannArrangement3_1_15_5}
	\end{center}
\end{figure}

In Fig. \ref{UllmannArrangement3_1_15_5} (a), we show one of the islands of the mode $(3, 1)$. Analogous to the previous arrangement, bifurcations eventually occur within islands of the mode $(3, 1)$ when $\epsilon_2$ is varied. The corresponding winding number profile for this initial panel is monotonic, as it can be seen in Fig.e (\ref{UllmannArrangement3_1_15_5}) (b). As $\epsilon_2$ increases,  the island is deformed and no bifurcations can be observed. It should be noted that all the winding number profiles depicted in Fig. (\ref{UllmannArrangement3_1_15_5}) are determined by the line segment connecting elliptic point of the island of mode $(3, 1)$ to its separatrix.

We note shearless curves emerging, highlighted in red, in the Poincaré sections of Fig. (\ref{UllmannArrangement3_1_15_5}). In Fig. (\ref{UllmannArrangement3_1_15_5}) (d), appears two local extremes in the winding number profile, corresponding to the first pair of shearless curves shown in Fig. (\ref{UllmannArrangement3_1_15_5}) (c). In addition to the emergence of local extremes and the transformation of the profile into a nonmonotonic form, the profile itself has exhibited some degree of rippling. This phenomenon can be observed in the interval $[0.155,0.165]$ for values of $y$ in Fig. (\ref{UllmannArrangement3_1_15_5}) (d).

In the profile depicted in Fig. \ref{UllmannArrangement3_1_15_5} (f), when $\epsilon_2 = 0.004519$, a second pair of extremes emerges within the interval where ripples were previously observed. The aforementioned extremes are initially in close proximity, as evidenced by their corresponding profile and the related Poincaré sections, in Fig. (\ref{UllmannArrangement3_1_15_5}) (e). Nevertheless, this pair of extremes (and shearless curves) will become more pronounced as $\epsilon_2$ increases.

In Fig. \ref{UllmannArrangement3_1_15_5} (h), two additional pairs are observed in the same region. It seems that the rippled in the winding number profile announces the emerging of extremes points. Fig. \ref{UllmannArrangement3_1_15_5} (g) depicts the island under investigation, which is now populated by many shearless curves. In total, there are eight shearless curves, with four related to local maximum in the winding number profile and four related to local minimum.

As the value of $\epsilon_2$ continues to increase, the emergence of internal island bifurcations can be observed, which ultimately contribute to the formation of mode $(15, 5)$. All shearless curves determined by local minimum in the profiles will become separatrixes, and no further new shearless curves will be generated. This configuration of $\{(m_1, n_1), (m_2, n_2)\}$ illustrates the scenario in which the shearless curves appear in pairs and allows us to designate one pair of these curves for each bifurcation.

\subsection{Emergence of shearless curves in different islands}

\begin{figure}[H]
	\begin{center}\includegraphics[width=0.6\textwidth]{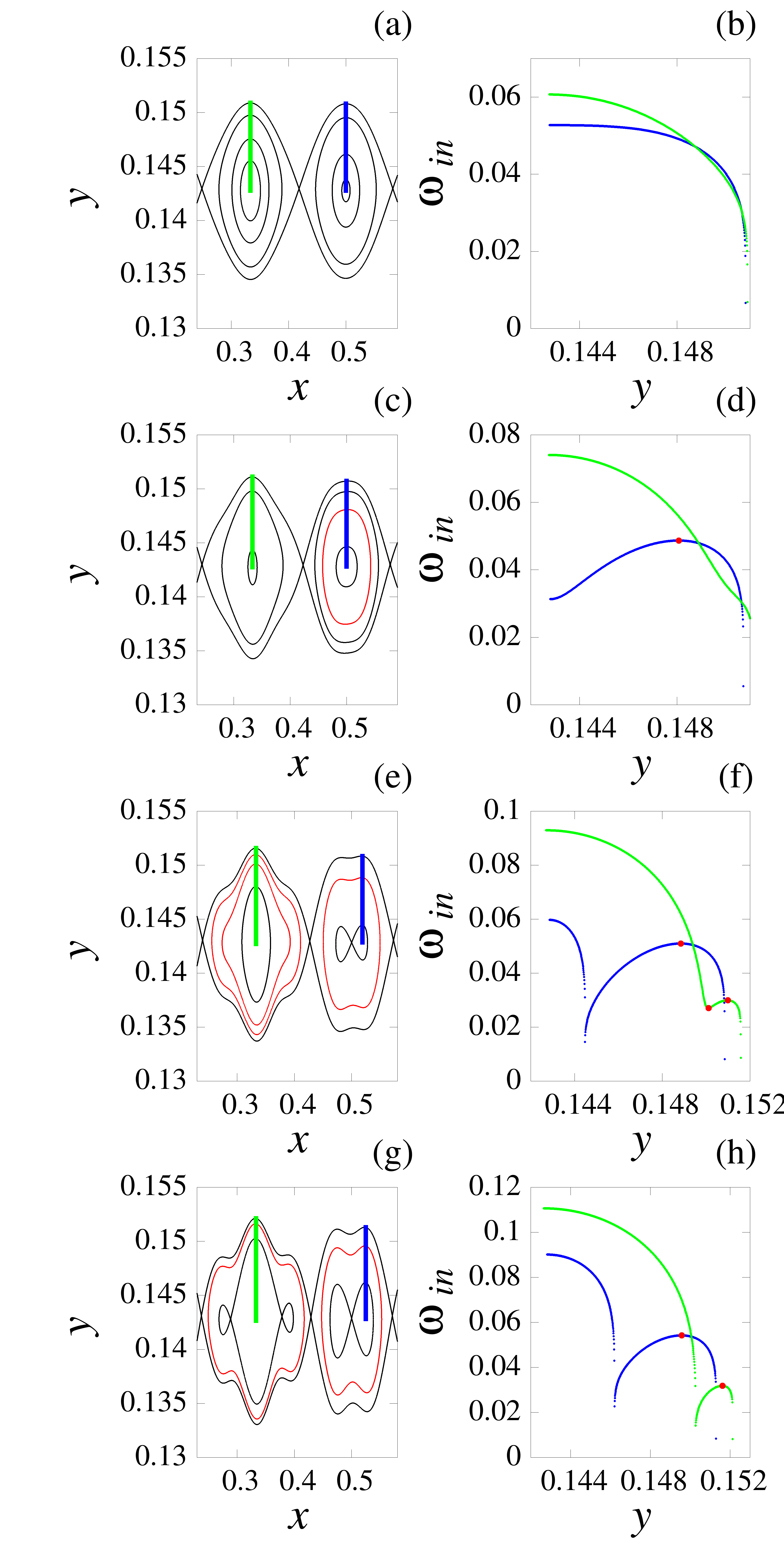}		
	\caption{Arrangement $\{(6, 2), (15, 5)\}$. (a) $\epsilon_2 = 0.000113\,(10A)$, (c) $\epsilon_2 = 0.000565\,(50A)$, (e) $\epsilon_2 = 0.001356\,(120A)$ and (g) $\epsilon_2 = 0.002259(200\,A)$. For all cases, $\epsilon_1 = 0.001130\,(100A)$.}		\label{ullmannArrangement6_2_15_5}
	\end{center}
\end{figure}

The final case presented using the Ullmann map displays behaviors that are similar to those observed in the previous cases, but simultaneously. The investigation commences at the mode $(6, 2)$, which is a mode formed by two isochronous islands $(3, 1)$. This is the reason we present two islands simultaneously in Fig. \ref{ullmannArrangement6_2_15_5} (a), \ref{ullmannArrangement6_2_15_5} (c), \ref{ullmannArrangement6_2_15_5} (e) and \ref{ullmannArrangement6_2_15_5} (g). Each of these islands is a member of one of the two $(3, 1)$ island chain. In the aforementioned four panels, we have maintained $\epsilon_1 = 0.001130$ and increased $\epsilon_2$.

In Fig. \ref{ullmannArrangement6_2_15_5} (a), the island situated to the leftmost exhibits internal winding number profile in green color, as illustrated in Fig. \ref{ullmannArrangement6_2_15_5} (b). While the island furthest to the right has internal winding number depicted in blue color. Both profiles are monotonic and $\epsilon_2 = 0.000113$.

When $\epsilon_2 = 0.000565$, while the profile related to the green color remains monotonic, the profile  in blue, that was monotonic in Fig. \ref{ullmannArrangement6_2_15_5} (b), has turned nonmonotic as shown in Fig. \ref{ullmannArrangement6_2_15_5} (d). This change is accompanied by the appearance of one shearless curve, as depicted in Fig. \ref{ullmannArrangement6_2_15_5} (c) by the red color curve on the largest island furthest to the right. It is worthy of note that the bifurcation inside this island is identical to the one showed in Fig. \ref{arrangment3_1_12_4} (b).

Eventually, the green profile also exhibits a nonmonotonic profile. As illustrated in Fig. \ref{ullmannArrangement6_2_15_5} (f), there are two local extremes, one local minimum and one local maximum, corresponding to the pair of shearless curve\textcolor{blue}{s} in Fig. \ref{ullmannArrangement6_2_15_5} (e), for $\epsilon_2 = 0.001356$.

In this example, the bifurcation is similar to the one shown in Fig. (\ref{fig3}) (c). It should be noted, however, that the emergence of the shearless curves is not associated with a double saddle-node bifurcation. Instead, two saddle-node bifurcations occur for values that are very close to each other (the results shown in reference\cite{leal2024} suggest that the bifurcations in the Ullmann map appears one by one). Nevertheless, the final number of shearless curve is identical to that observed in the double bifurcation.

Lastly, for $\epsilon_2 = 0.002259$, the shearless curve determined by the local minimum in Fig. \ref{ullmannArrangement6_2_15_5} (f), the green profile, becomes an internal separatrix of the largest island to the left in Fig. \ref{ullmannArrangement6_2_15_5} (g). The winding number profiles associated to this Poincaré section are illustrated in Fig. \ref{ullmannArrangement6_2_15_5} (h). It can be observed that the shearless curve by the local maximum survives as $\epsilon_2$ increased.

\section{Walker-Ford Hamiltonian}
\label{sec:WF}

Complementary to the discrete maps shown so far, we also included a Hamiltonian flow as an alternative framework for the bifurcations and the emergence of shearless invariants. The Hamiltonian model in question was initially meant to describe the two-dimensional dynamics of a star moving around a galaxy center with a cylindrical symmetric potential. It was first used to numerically prove the emergence of chaos due to the lack of a third integral of motion when perturbations acted on the system \cite{Henon}.  

In action-angle format, its non-perturbed (hence integrable) form reads
\begin{equation}\label{eq:walker-ford-non-perturbed}
    H_0(J_1, J_2) = J_1 + J_2 - J_1^2 - 3 J_1 J_2 + J_2^2, 
\end{equation}
for $J_1$ and $J_2$ as action variables.
From this point, Walker and Ford consider two coupled oscillators to simulate the effects of single and double resonances on the appearance of stability islands in phase space \cite{Walker-Ford,Sousa}. Thus, the complete Hamiltonian considers the non-perturbed term $H_0$ and two controlled perturbations
\begin{equation}\label{eq:walker-ford-triple-mode}
    H(\theta_1, \theta_2, J_1, J_2) = H_0 + H_1 + H_2,
\end{equation}
with each perturbation mode as
\begin{equation}\label{eq:walker-ford-modes-list}
\begin{split}    
    H_1(\theta_1, \theta_2, J_1, J_2) &= \alpha J_1 J_2 \cos(m\theta_1 - n\theta_2) \\
    H_2(\theta_1, \theta_2, J_1, J_2) &= \beta  J_1 J_2^{\frac{3}{2}} \cos(r\theta_1 - s\theta_2),
\end{split}
\end{equation}
where $(m,n)$ and $(r,s)$ are the modes controlling the amount of islands generated by $H_1$ and $H_2$, respectively. The amplitudes $\alpha$ and $\beta$ control their size.

The phase space analysis in this case takes place for a discrete map generated by the intersection of the flow trajectories with a surface section -- here taken as $(\theta_1 = \frac{3 \pi}{2}$). In general, we still have a resonance between modes, although now corresponding to oscillatory terms in the energy function (\textit{i.e.} the Hamiltonian itself in the present model).

For the purpose of this work, we fixed the amplitude and mode of $H_1$ to $\alpha=0.02$ and $(m,n)=(2,2)$, focusing on the effect of the second perturbation $H_2$ as its amplitude and modes change. 
This limitation of equal mode numbers is meant to reduce the dynamics in phase space, so as to isolate and reproduce the same scenarios seen for the two-harmonic standard map and Ullmann's map. Moreover, in this arrangement, since resonance modes are multiple of each other, no chaos is generated, as it is always possible to define a single angle variable $\Theta = \theta_1 - \theta_2$, reducing Hamiltonian (\ref{eq:walker-ford-triple-mode}) to a one-degree-of-freedom system with $H$ as constant of motion (given its explicit time independence). However, non-multiple modes are enough to induce chaos in the system if desired.

\subsection{Emergence of one shearless curve}

\begin{figure}[H]
    \centering
    \includegraphics[width=1.0\textwidth]{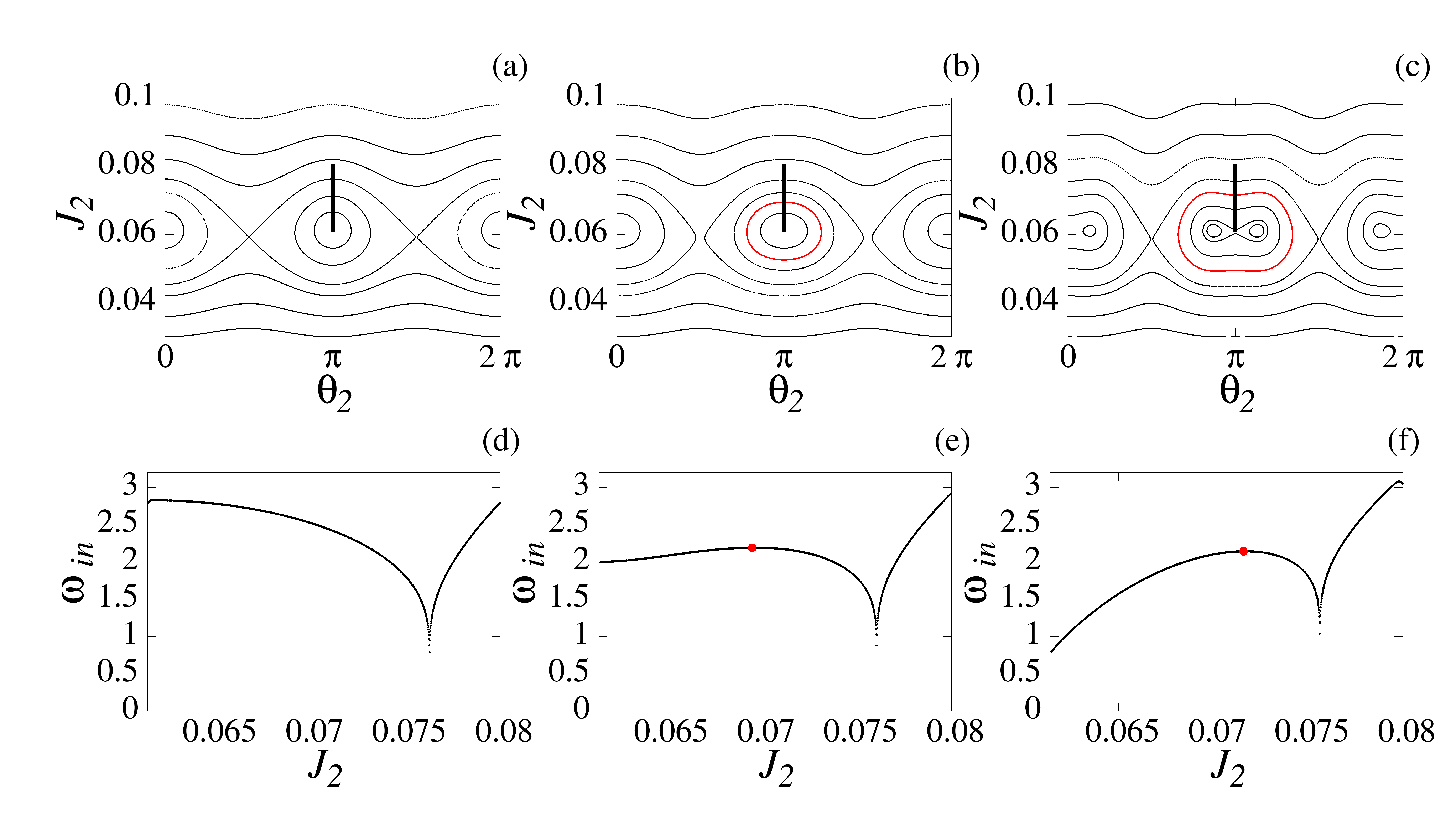}
    \caption{Arrangement $\{(2,2), (4,4)\}$. Panels (a), (b) and (c) show $\beta=0.0$, $0.01$ and $0.02$, respectively. The corresponding winding number profile $\omega_{in}$ is shown below each frame, \textit{i.e.} (d), (e) and (f). In all cases, $\alpha=0.02$, $H=0.22$. }
    \label{fig:mn2_rs4}
\end{figure}

Primarily, in the absence of the second perturbative mode (when $\beta=0$), phase space presents isolated islands in a period-two chain, as generated by the $(m,n)$ mode (Fig. \ref{fig:mn2_rs4} (a)) with monotonic winding number profile (Fig. \ref{fig:mn2_rs4} (d)). As soon as the second mode is introduced, both islands maintain their internal structure without bifurcation for $\beta < 0.02$  (Fig. \ref{fig:mn2_rs4} (b)) although with a winding profile presenting a slight plateau (Fig. \ref{fig:mn2_rs4} (e)).

For further increasing $\beta \geq 0.02$, the second mode resonates with the first and induces a pitchfork bifurcation in both island centers independently, thereby generating a period-two chain with inner resonances also of period two (Fig. \ref{fig:mn2_rs4} (c)). These resonances increase in size with $\beta$, but without further bifurcations or emergence of extremant points in the rotation number profile. 
Despite inducing both phenomena, the shearless curve and the bifurcation, they are not simultaneous in this simplest case. This scenario corresponds to the same seen in Fig. \ref{fig1}. 

\subsection{Emergence of multiple shearless curves}

\begin{figure}[H]
    \centering
    \includegraphics[width=1.0\textwidth]{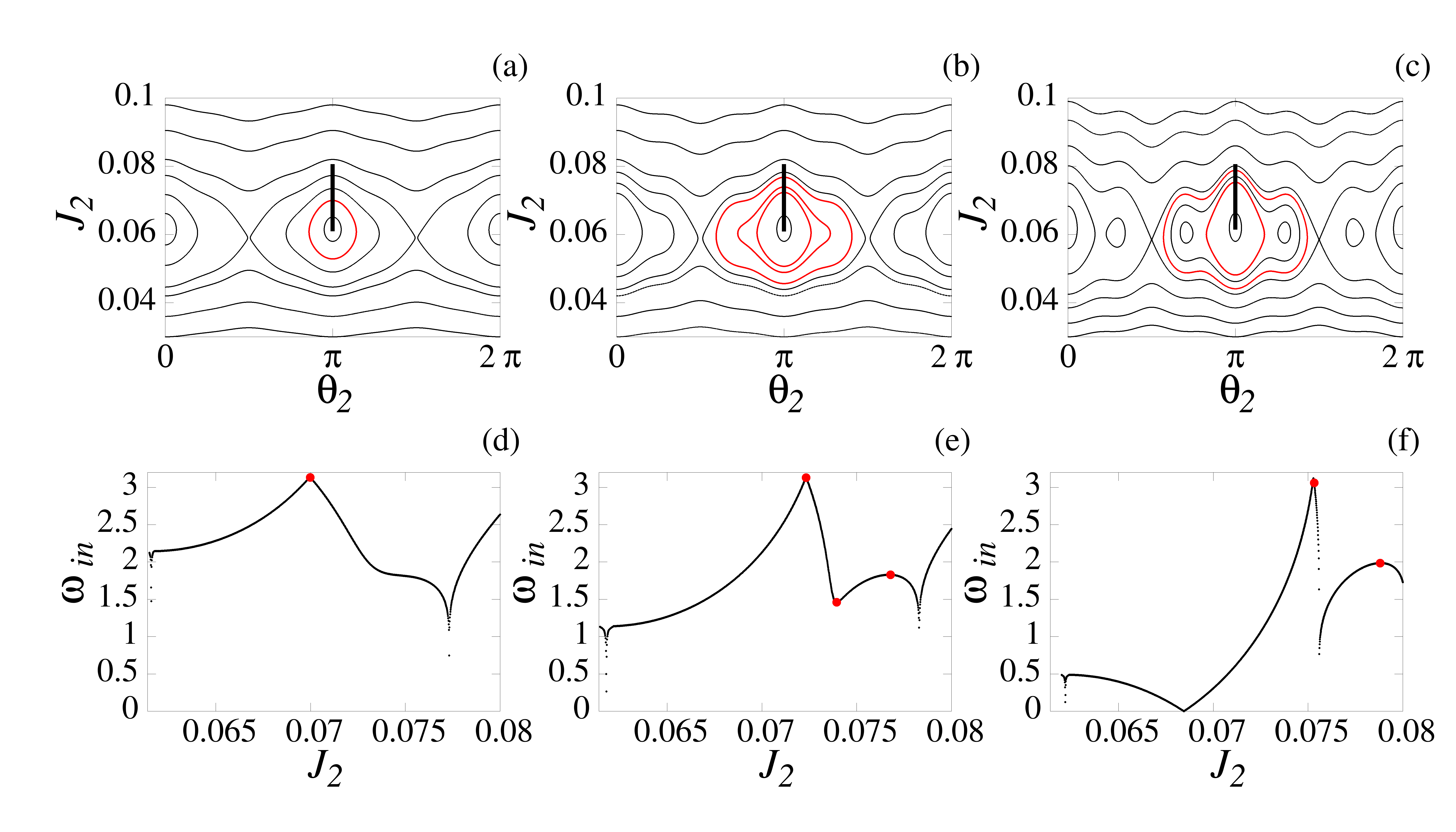}
    \caption{Arrangement $\{(2,2), (6,6)\}$. Panels (a), (b) and (c) show $\beta=0.01, 0.02$ and $0.04$, respectively. The corresponding winding number profile $\omega_{in}$ is shown below each frame, \textit{i.e.} (d), (e) and (f). In all cases, $\alpha=0.02$, $H=0.22$. }
    \label{fig:mn2_rs6}
\end{figure}

When changing the perturbation mode to $r=s=6$, for low amplitude ($\beta=0.01$ -- Fig. \ref{fig:mn2_rs6} (a)) the island structure is still similar to cases (a) and (b) for $r=s=4$, but with a winding profile presenting a maximum point and two inflections (near $J_2=0.075$ (Fig. \ref{fig:mn2_rs6} (d)).
At slightly higher $\beta = 0.02$, these inflection points shift and become a mininum-maximum pair (Fig. \ref{fig:mn2_rs6} (e)), therefore creating two extra shearless curves within the central island (Fig. \ref{fig:mn2_rs6} (b)), although with no bifurcation occurring. 

This process is roughly the same seen from Fig. \ref{fig2} (d) to \ref{fig2} (e) and \ref{fig2} (e) to \ref{fig2} (f),  Fig. \ref{fig3} (d) to \ref{fig3}(e) and in Fig. \ref{arrangment3_1_12_4} (e) to (f) .
At $\beta = 0.04$, the previous minimum point pinch down in the winding profile (Fig. \ref{fig:mn2_rs6} (f)), forming a pit. In phase space, the previously shearless curve related to this minimum becomes the separatrix of the new islands bifurcated around the central elliptical point.
The two previous maxima remain, each one forming an inner and an outer shearless invariant (Fig. \ref{fig:mn2_rs6} (c)).

The comparison between the simplest mode $r=s=4$ with $r=s=6$ suggests that the number of extremes in the winding profile grows with the period of the second perturbation and therewith the presence of shearless invariants.

\subsection{Emergence of shearless curves in different islands}

At last, the two scenarios shown so far correspond to resonance parameters $m,n,r$ and $s$ as multiples and even numbers. This implies that resonances promote bifurcations near or over the elliptical central points, as they have similar parity.
To complement these results, we also present the bifurcation behavior for an odd perturbation mode, that is $r=s=5$.

\begin{figure}[H]
    \centering
    \includegraphics[width=0.8\textwidth]{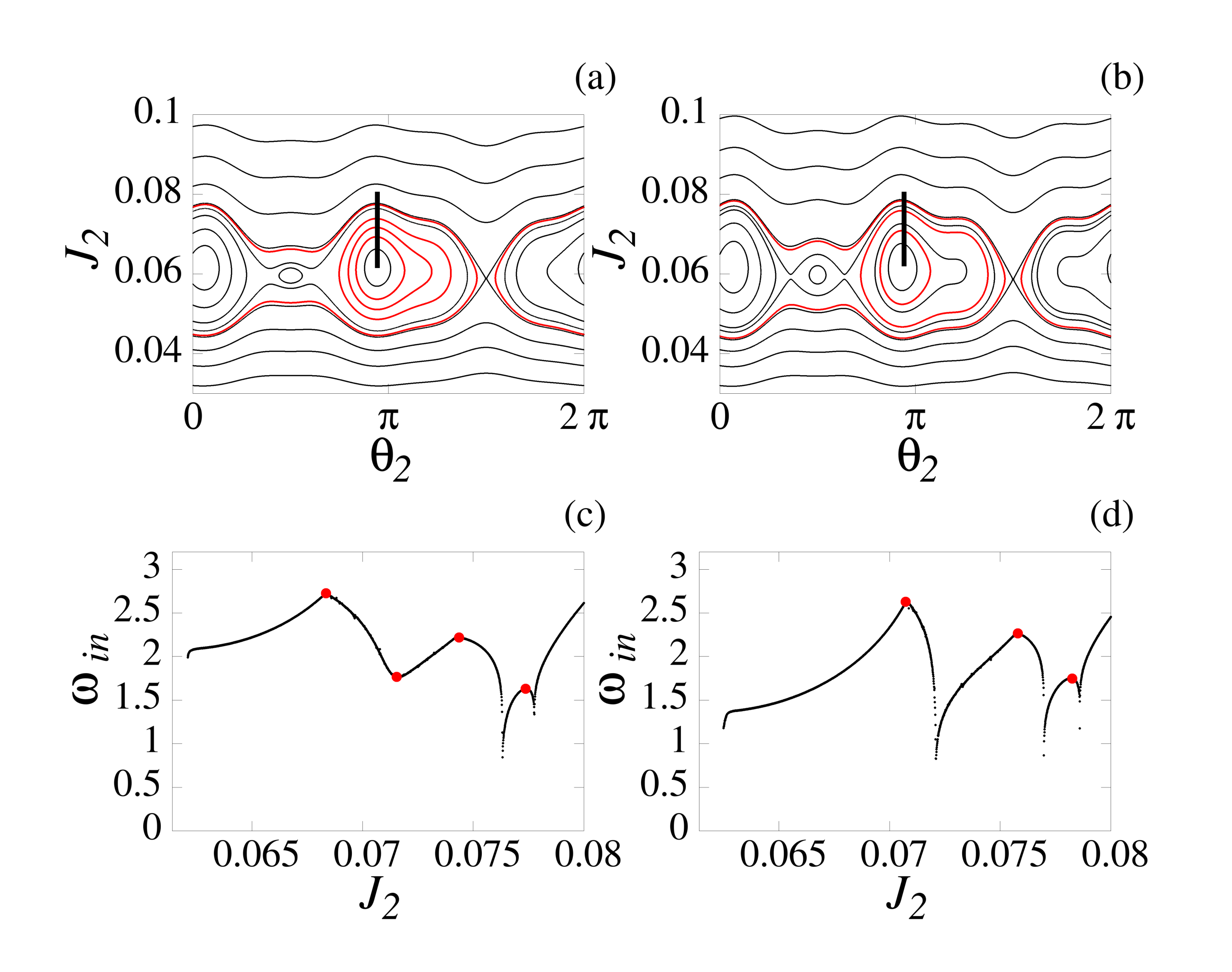}
    \caption{Arrangement $\{(2,2), (5,5)\}$. Panels (a) and (b) show $\beta=0.02$ and $0.03$, respectively. The corresponding winding number profile $\omega_{in}$ is shown below each frame, \textit{i.e.} (c) and (d). In all cases, $\alpha=0.02$, $H=0.22$.}
    \label{fig:mn2_rs5}
\end{figure}

As shown in Fig. \ref{fig:mn2_rs5}, most features are similar to what was seen for even $(r,s)$, with the minimum point at $J_2 = 0.0715$ pinching down, while its corresponding shearless invariant becomes the separatrix of the new side island at its right (Fig. \ref{fig:mn2_rs5} (b) and \ref{fig:mn2_rs5} (d)). On the other hand, given the distribution of the bifurcated islands, with elliptical center at $\theta_2=\pi/2$ and side island at $\theta_2\approx \pi+\pi/10$, the shearless curve extends over all the island chain, whereas the inner one remains within the previous central island.

\section{Conclusions}
\label{sec:Conclusions}

The onset of secondary shearless curves has been previously  reported in a few twist systems. In this article we present further examples where we identify the presence of such curves in different coupled  systems. Our results indicate that secondary shearless curves are commonly observed in twist systems.

By employing the two-harmonic standard map, a conservative map, we were able to discern three patterns for the emergence of shearless curves. As the perturbation parameter increases, a single shearless curve, defined by a local maximum point in the internal rotation number profile, emerges and characterizes one of the patterns. A second pattern is the emergence of shearless curves in pairs. These curves manifest as maximum and minimum pairs in the internal rotation profile. Ultimately, the shearless curve defined by the local minimum point will evolve into a separatrix in scenarios where the bifurcations are of the saddle-node type. Finally, a third pattern is observed, pairs of shearless curves precede a double saddle-node bifurcation. In this instance, two simultaneous saddle-node bifurcations result in a single pair of shearless curves.

Alternatively to the standard map, we made use of another conservative map which describes magnetic field lines for confinement of plasma in toroidal geometries. For this system, we were also able to identify the three patterns of emergence of shearless curves discussed in the previous paragraph. The only difference being that shearless curves precede only saddle-nodes and they emerge one by one.

In addition to discrete systems, we use a Hamiltonian flow, specifically the Walker-Ford Hamiltonian, to show the emergence of shearless curves in pairs, or solely as observed on the previously studied maps.  Preceding a pitchfork type bifurcation, a single shearless curve appears,  as defined by the emergence of a maximum in the rotation number profile. Furthermore, saddle-node bifurcations  give rise to pairs of shearless curves; in accordance to the pairs of maxima and minima in the rotation profile. Nevertheless, for the Hamiltonian flow, new phenomenons were also observed, when compared to the discrete maps. The emergence of a shearless curve apparently without the subsequent appearance of a bifurcation and the inversion of the local maximum by a minimum in the winding number profile together with the appearance of a new pair of shearless curves.

In summary, we numerically identified the occurrence of shearless curves in discrete and continuous twist Hamiltonian systems with coupling of resonant modes. We noticed that the curves can appear either alone, or in pairs, and always preceding fixed point bifurcations. When appearing in maximum-minimum pairs, the curves determined by the local minimum in the rotation profile eventually become separatrices.
\section*{Acknowledgments} 

This research received the support of the Coordination for the Improvement of Higher Education Personnel (CAPES) under Grant No. 88887.320059/2019-00, 88881.143103/2017-01, the National Council for Scientific and Technological Development (CNPq - Grant No. 403120/2021-7, 311168/2020-5, 301019/2019-3), Fundação de Amparo à Pesquisa do Estado de São Paulo (FAPESP) under Grant No. 2022/12736-0, 2018/03211-6, 2023/10521-0 and CNEN (Comissão Nacional de Energia Nuclear) under Grant No. 01341.001299/2021-54.

\section{Appendix}
\label{apend}

In the context of magnetic confinement plasma, with the ultimate goal of achieving power fusion, there is a number of confinement settings that employ a magnetic perturbation. The primary objective of such perturbation is to generate a chaotic magnetic layer that can mitigate certain types of instabilities or allow the escape of the magnetic lines in a suitable manner. Some examples of these devices are the divertor\cite{wesson2004book}, saddle coils as in DIII-D \cite{abdullaev2014magnetic}, and the magnetic ergodic limiter\cite{pires2005}.

We have a symplectic mapping \cite{ott2002book} {given by equations (\ref{UllmannEqulRadial}), (\ref{UllmannEqulPoloidal}) and (\ref{ullmannPertRad}), (\ref{ullmannPertPol})} that describes the evolution of a magnetic field line configuration capable of confining a plasma along a torus. The model, as determined by the mapping, adopts a magnetic field of equilibrium $\mathbf{B_{eq}}$ (composition of a poloidal and a toroidal magnetic fields) that is periodically perturbed by an ergodic magnetic limiter, which is essentially  a set of coils through which electric currents flow.

It is important to note that the periodicity occurs in a geometrical sense, not in a time sense; all the confinement is static. The mapping was originally proposed by Ullmann \cite{ullmann2000}, while the perturbation setup is analogous to that proposed in B.B.Leal \textit{et al}\cite{leal2024}. 

This mapping employs coordinates radial $r$ and poloidal $\theta$, not action and angle variables, to describe the position of a magnetic field line. They are analogous to cylindrical coordinates and since we study the dynamics by a plane crossing transversally the torus, there is no need for the use of toroidal coordinate $\phi$. By construction, the Poincaré section must be localized adjacently to one of the coils. As discussed before, the equations that describe the magnetic field  line positions, at the plasma confinement equilibrium, are given by (\ref{UllmannEqulRadial}) and (\ref{UllmannEqulPoloidal}).

The equations (\ref{UllmannEqulRadial}) and (\ref{UllmannEqulPoloidal}) determine the toroidal evolution of a magnetic field line that has been previously localized in $(r, \theta)$ to the position $(r^*, \theta^*)$, after $2 \pi / N$ rad toroidally shift. The magnetic field line at $(r^*, \theta^*)$ is always the position after a toroidally evolution of $2 \pi / N$ rad and before the perturbation due to the ergodic magnetic limiter and $(r, \theta)$ is always the magnetic position line after the perturbation. For convenience, we adopted a cylindrical approximation  because it reduces the emerging of chaotic field lines. Therefore, the torus is approximated by a periodic cylinder of length $2 \pi R$ and the equations (\ref{UllmannEqulRadial}) and (\ref{UllmannEqulPoloidal}) can be deduced from $\mathbf{B_{eq.}} \times {d}\mathbf{l} = {\mathbf{0}}$, where ${d}\mathbf{l}$ is an infinitesimal segment of the filed line. 

{In equation (\ref{UllmannEqulPoloidal}) appears $q(r)$, the safety factor, an} important measure at plasmas confinement in tokamaks. For example, a tokamak must operate with values of $q>1$ to avoid kink mode instabilities. It is defined as the mean value of the ratio of the toroidal variation $\Delta \phi$ to the poloidal variation $\Delta \theta$, $\nu \equiv \Delta \phi / \Delta \theta$,
\begin{equation}
    q \equiv \frac{1}{2 \pi} \int{\nu d \theta},
\end{equation}
and it has a simple relation with the winding number. One is the inverse of the other $\omega = 1 / q$.

\begin{figure}[!h]
    \centering
    \includegraphics[width=0.35\textwidth]{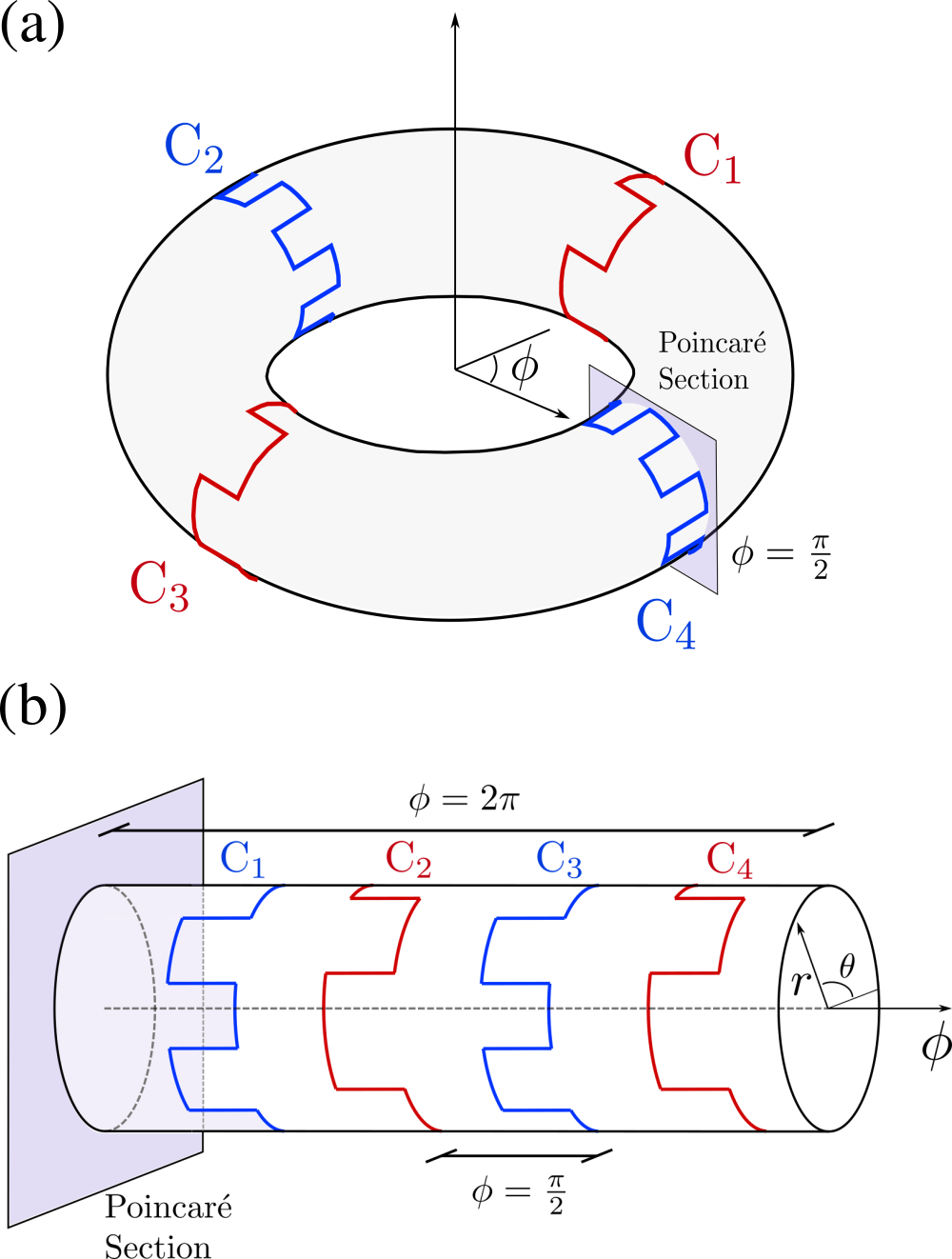}
    \caption{a) The magnetic coils along the toroidal chamber are equally separated from each other. In b), the cylindrical approach is illustrated.  The toroid is viewed as a $2 \pi$R-periodic cylinder.}
    \label{illustrationTokamak}
\end{figure}

{The magnetic field line equilibrium is perturbed by a magnetic field, whose mapping is given by equations (\ref{ullmannPertRad}), (\ref{UllmannEqulPoloidal}). The perturbation is  generated by a set of coils called ergodic magnetic limiter.} These coils produce resonances on magnetic field line surfaces with a rational internal winding number. The set of coils is formed by two pairs, each of which must produce a mode perturbation of type $(m, n)$. The distance between each coil of one pair is equal to $\pi$ radians, and the toroidal displacement between a given coil and its neighbor is $\pi / 2$ rad. In Fig.(\ref{illustrationTokamak} (a) there is an illustration of how the magnetic coils are distributed along the toroidal chamber, while Fig.(\ref{illustrationTokamak} (b) represents the cylindrical approach. 

As previously mentioned, each pair $i$ of coils generates a resonance of type $(m_i , n_i)$. In order to achieve this, it is necessarily for each coil to have a poloidal twist in relation to its counterpart coil. Consequently, the parameter $\alpha_i = \pi \frac{n_i}{mi}$ must be included as a phase in the trigonometric functions of the perturbation map (\ref{ullmannPertRad}, \ref{ullmannPertPol}).

The equations (\ref{ullmannPertRad}) and (\ref{ullmannPertPol}) are the same used in the work\cite{leal2024} and is a version of the Ullmann map \cite{ullmann2000}. The parameter $b$ is the minor radius of the tokamak, $m$, the poloidal mode, is the number of pairs of toroidal wires segments\cite{pires2005} (it produces chains of $m$ islands).
\label{apendA}

\end{document}